\documentclass[floatfix,aps,prl,reprint,groupedaddress,amsmath,amssymby,longbibliography]{revtex4-1}
\usepackage[english]{babel}
\PassOptionsToPackage{english}{babel}

\usepackage[utf8]{inputenc}
\usepackage{color}
\usepackage{amsmath,amssymb,graphicx,xcolor,units,hyperref}

\newcommand{\dd}[1]{\mathrm{d}#1\,}

\renewcommand{\Re}{\mathop{\mathrm{Re}}}
\renewcommand{\Im}{\mathop{\mathrm{Im}}}

\DeclareMathOperator{\artanh}{artanh}

\begin{document}
\selectlanguage{english}
\title{Phase-driven collapse of the Cooper condensate in a nanosized superconductor}

\author{Alberto Ronzani}
\email{alberto.ronzani@sns.it}
\affiliation{NEST, Istituto Nanoscienze-CNR and Scuola Normale Superiore, I-56127 Pisa, Italy}
		                  
\author{Carles Altimiras}
\affiliation{SPEC, CEA, CNRS, Universit\'e Paris-Saclay, CEA-Saclay, 91191 Gif-sur-Yvette, France}

\author{Sophie D'Ambrosio}
\affiliation{NEST, Istituto Nanoscienze-CNR and Scuola Normale Superiore, I-56127 Pisa, Italy}

\author{Pauli Virtanen}
\affiliation{NEST, Istituto Nanoscienze-CNR and Scuola Normale Superiore, I-56127 Pisa, Italy}

\author{Francesco Giazotto}
\email{francesco.giazotto@sns.it}
\affiliation{NEST, Istituto Nanoscienze-CNR and Scuola Normale Superiore, I-56127 Pisa, Italy}

\maketitle

\bfseries
Superconductivity can be understood in terms of a phase transition from an
uncorrelated electron gas to a condensate of Cooper pairs in which the relative
phases of the constituent electrons are coherent over macroscopic length
scales~\cite{de_gennes_superconductivity_1999}.
The degree of correlation is quantified by a complex-valued order parameter,
whose amplitude is proportional to the strength of the pairing potential in 
the condensate. Supercurrent-carrying states are
associated with non-zero values of the spatial gradient of the phase.
The pairing potential and several physical observables of the
Cooper condensate can be manipulated by means of temperature,
current bias, dishomogeneities in the chemical composition or application of
a magnetic field~\cite{tinkham_introduction_1996}.
Here we show evidence of complete suppression of the energy gap in the local
density of quasiparticle states (DOS) of a superconducting nanowire upon establishing a
phase difference equal to $\pi$ over a length scale comparable to the
superconducting coherence length.
These observations are consistent with a complete collapse of the pairing
potential in the center of the wire, in accordance with theoretical modeling based 
on the quasiclassical theory of superconductivity in diffusive
systems.
Our spectroscopic data, fully exploring the phase-biased states of the
condensate, highlight the profound effect that extreme phase gradients
exert on the amplitude of the pairing potential.
Moreover, the sharp magnetic response observed near the onset of the
superconducting gap collapse regime can be exploited to realize ultra-low noise
magnetic flux detectors~\cite{ronzani_highly_2014}.

\normalfont

\begin{figure}[t!]
	\centering
	\includegraphics[width=0.45\textwidth]{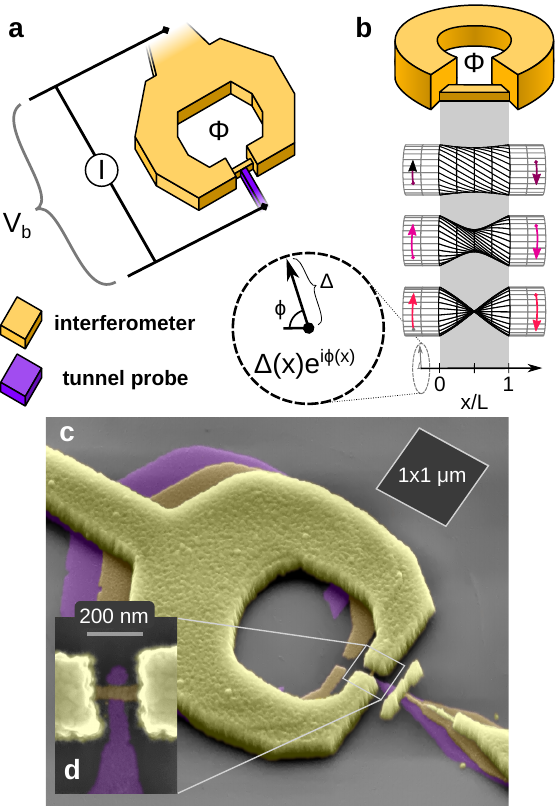}
	\caption{Principle of operation and interferometer design.
		\textbf{a}: Schematic representation of the measurement
	setup for the transport spectroscopy of a phase-biased superconducting
	wire. The current-voltage characteristics show the modulation of the
	local density of states in the latter as a function of the magnetic flux
	$\Phi$ coupled to a compact superconducting loop in clean contact with the wire.
	\textbf{b}: Conceptual picture of the progressive depairing in
	the middle of the superconducting wire for increasing phase gradient.
	The position-dependent value of the complex order parameter $\Delta(x)
	\exp\left[\,\mathrm{i} \phi (x)\right]$ is shown as a twisted-wireframe
	representation of a revolution surface. For a superconducting wire
	having single-valued current-to-phase relationship, $\Delta(L/2) = 0$ is
	expected for $\Phi = \Phi_0/2$.
	\textbf{c, d}: Pseudo-colour scanning electron micrographs of, respectively, 
	the interferometer loop and the superconducting wire in a typical device.
	Here, yellow indicates the $150$-nm-thick
	``interferometer'' and $25$-nm-thick ``nanowire'' Al layers; the
	$15$-nm-thick AlOx tunnel probe electrode is shown in purple,
	realized by oxidizing respectively a Al/Al$_{0.98}$Mn$_{0.02}$
	layer to obtain a superconducting/normal-metal electrode.
	}\label{fig:concept}
\end{figure}

The physics of superconducting boundaries
is an invaluable tool for the investigation of the fundamental
properties of matter and for fostering the development of novel devices. From
tunnel-type contacts~\cite{giaever_electron_1960} to clean galvanic interfaces
dominated by proximity effect~\cite{de_gennes_superconductivity_1999},
the fabrication and
manipulation of superconducting boundaries is the enabling feature for the
realization of ultrasensitive magnetometers~\cite{vasyukov_scanning_2013} and
electrometers~\cite{schoelkopf_radio-frequency_1998}, sub-Kelvin electron
thermometers and
coolers~\cite{giazotto_opportunities_2006}, coherent mesoscopic heat current
controllers~\cite{giazotto_josephson_2012,martinez-perez_rectification_2015}, 
radiation detectors~\cite{day_broadband_2003}, parametric
amplifiers~\cite{ho_eom_wideband_2012,macklin_nearquantum-limited_2015},
qubits~\cite{clarke_superconducting_2008} and Majorana physics
demonstrators~\cite{mourik_signatures_2012}. 

\begin{figure*}[ht!]
	\centering
	\includegraphics{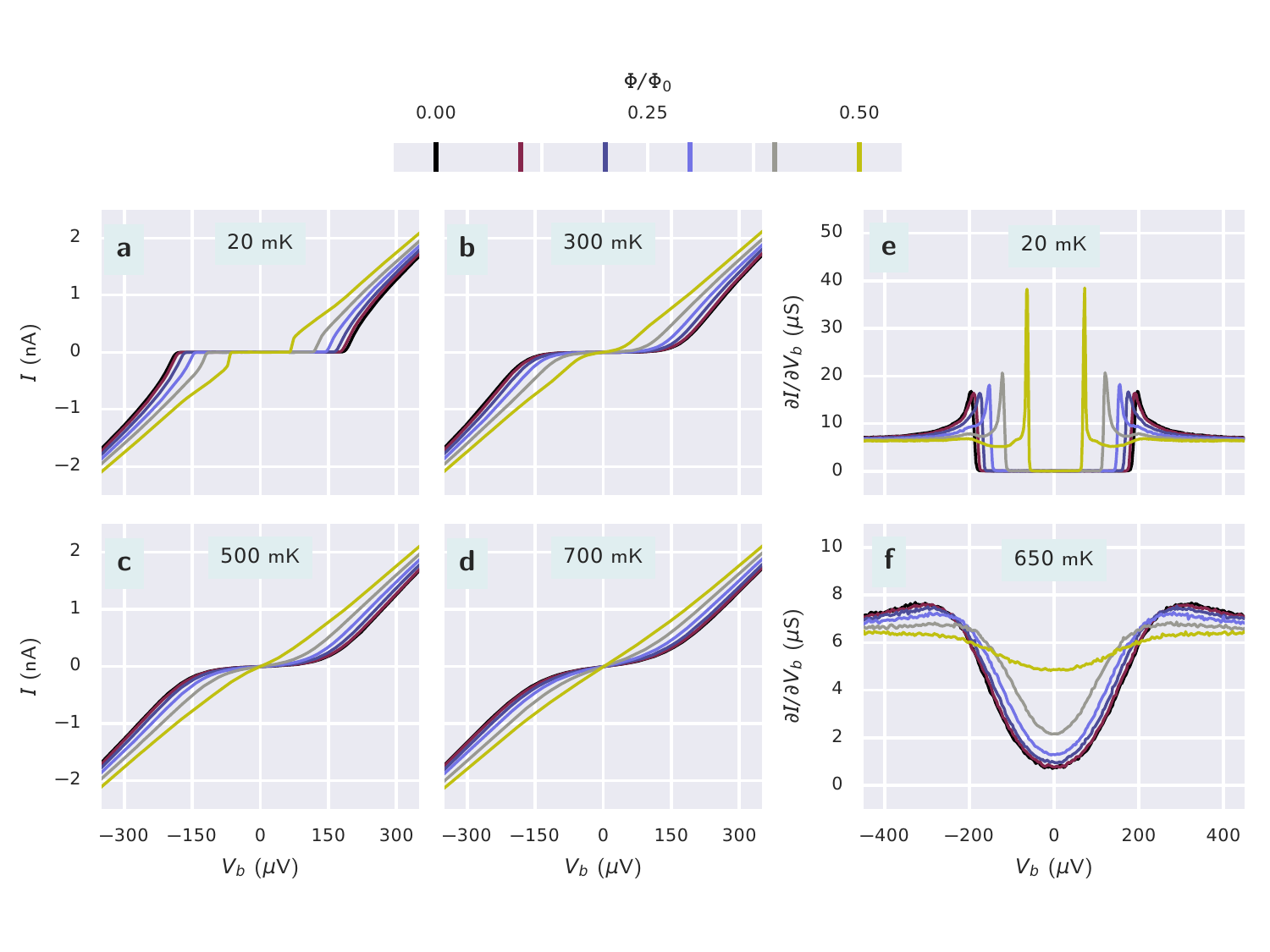}
\caption{Magneto-electric response of a typical normal-metal tunnel probe device.
	Different traces are color-coded to six applied magnetic flux values
	equally spaced in the range $\Phi = 0 \to \Phi_0/2$.
	\textbf{a--d}: 
	Current-vs-voltage characteristic curves, recorded 
	at lattice temperature $T=20,\, 300,\, 500,\, 700\, \mathrm{mK}$,
	respectively.
	\textbf{e, f}: Differential conductance as a function of voltage bias,
	recorded at lattice temperature $T=20,\, 650\, \mathrm{mK}$, respectively.
}
\label{fig:nprobe}
\end{figure*}

In superconducting electronics, localized geometrical and compositional
inhomogeneities provide
preferential pinning points for the establishment of order parameter phase
gradients, the distinctive superconducting degree of freedom.
This is exemplified by the coherent
character of the pair transport between two different Cooper condensates leading
to the Josephson effect~\cite{josephson_possible_1962}, 
where the current to phase relation (CPR) is determined
by the phase-dependent energy spectrum of weak link-bound
states~\cite{golubov_current-phase_2004}. 
The latter has been resolved spectroscopically down to the
individual mesoscopic channel in atomic contacts~\cite{bretheau_exciting_2013},
carbon nanotubes~\cite{pillet_andreev_2010} and semiconducting
nanowires~\cite{chang_tunneling_2013}.

\begin{figure*}[ht!]
	\centering
	\includegraphics{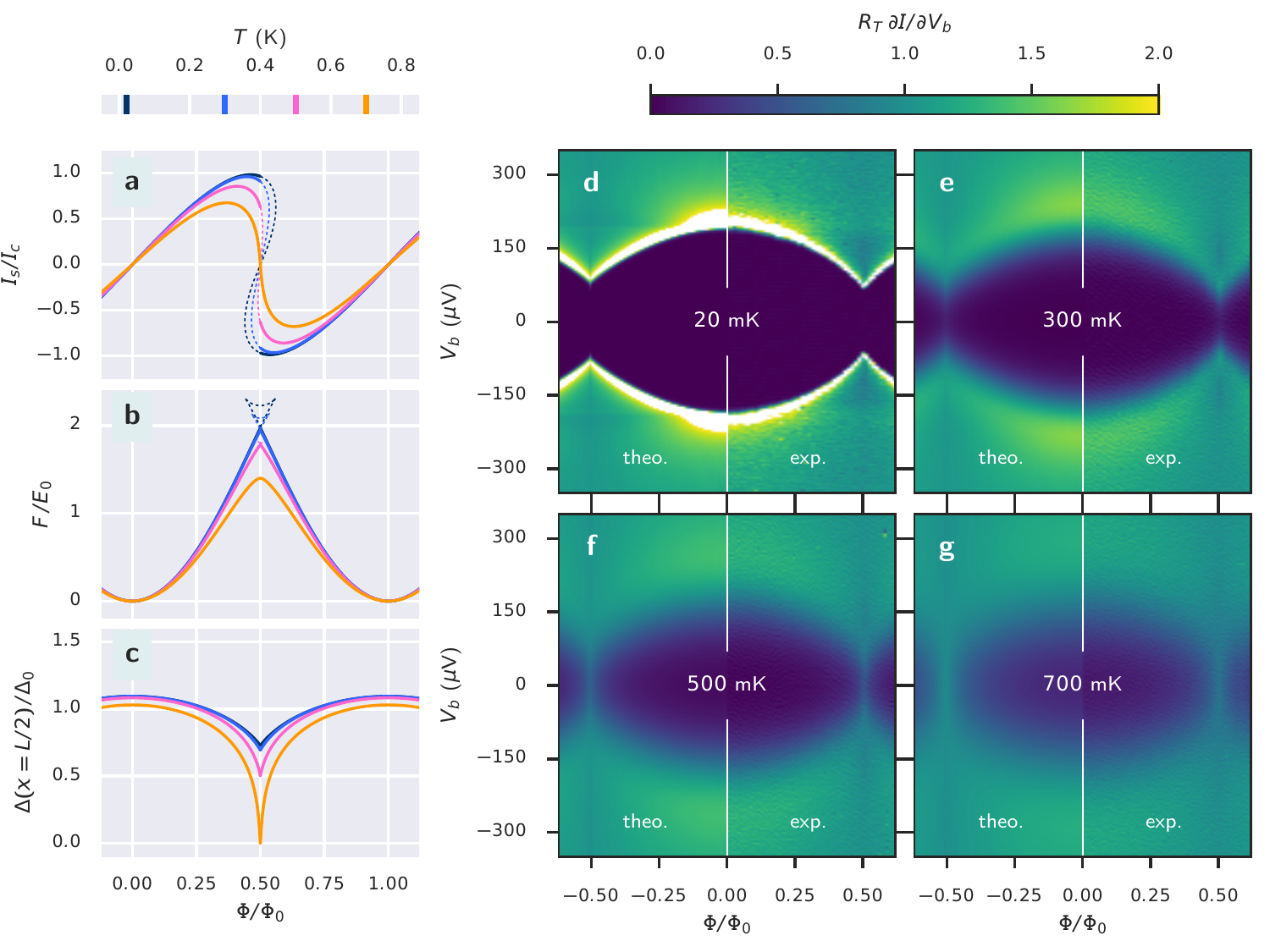}
\caption{Results for the modelization of the superconducting nanowire according
to parameters inferred from experimental data.
	\textbf{a}: Circulating supercurrent $I_s$ normalized to the
	zero-temperature value of the critical current of the wire ($I_c$).
	\textbf{b}: 
	Interferometer free energy $F$ normalized to $E_0 = I_c \Phi_0 / 2 \pi$.
	\textbf{c}: Pairing potential amplitude
	$\Delta(x=L/2)$ in the center of the superconducting nanowire 
	normalized to the zero-temperature value of the pairing potential
	amplitude in the interferometer loop ($\Delta_0$).
	In panels \textbf{a--c}, different traces are color-coded to the temperature values
	$T=20,\, 300,\, 500,\, 700\, \mathrm{mK}$ and show the modulation of the 
	respective quantity as a function of the magnetic flux $\Phi$ applied to
	the interferometer.  Dotted branches in panels \textbf{a, b} represent
	theoretical solutions corresponding to thermodynamically unstable
	interferometer states.
\textbf{d--g}: 
	Normalized differential conductance of the normal-metal tunnel probe
	device as a function of the applied magnetic flux and voltage bias $V_b$
	at lattice temperature $T=20,\, 300,\, 500,\, 700\, \mathrm{mK}$, 
	respectively.
	The halves of each colormap plot allow the
	comparison between theoretical predictions (left) and experimental data
	(right).
	}
\label{fig:theo}
\end{figure*}

Correspondingly, physical observables in diffusive mesoscopic weak links
are dependent on the phase difference between their superconducting
boundaries.  The phase-driven modulation of the DOS
of proximized normal metal weak links is a well known
example~\cite{le_sueur_phase_2008}.
Analogous observations have been reported for superconductive
wires in the long limit, where partial depairing was induced by spatially
uniform phase gradient profiles~\cite{anthore_density_2003}.

On the other hand, phase-biasing a weak link based on a thin superconducting
wire in the short limit (i.e., having comparable geometric and coherence
length) allows the observation of the reaction of
its Cooper condensate to a spatial phase profile well beyond the linear regime. 
In particular, provided that the state of the Cooper condensate is a
single-valued function of the phase difference applied to this weak link, the
latter can be polarized with a phase difference equal to $\pi$, reaching the
non-trivial node of its CPR. 
In this specific state no supercurrent flows in the Cooper
condensate and, by virtue of time-reversal symmetry, its order parameter
$\Delta \exp \left(\mathrm{i} \phi \right)$ is 
a real-valued and sign-changing function of the spatial
coordinate along the wire. Then, as a consequence of continuity, 
the pairing potential amplitude $\Delta(x)$ must equal zero in some position inside the
wire (e.g., the centre in case of symmetric boundaries).
While wide ($w \gtrsim \xi$) weak links can accommodate Abrikosov vortices,
quasi-1D wires ($w < \xi$) develop a 1D phase singularity without screening
currents~\cite{likharev_superconducting_1979}, namely a phase-slip center.
A conceptual representation of the mechanism of phase-driven collapse of the
order parameter is presented in Figure~\ref{fig:concept}b.

 In our experiment, the phase bias on a narrow nanosized aluminum wire
 is enforced by virtue of magnetic flux quantization in a closed
 superconducting loop subjected to a magnetic field applied
 orthogonally~\cite{doll_experimental_1961,deaver_experimental_1961}. 
 The phase-dependent DOS inside the wire is probed 
 through standard two-wire charge transport spectroscopy via a tunnel-barrier
 electrode (refer to the schematic in Figure~\ref{fig:concept}a). 
 This type of device is the superconducting-wire analogue of the
 superconducting quantum interference proximity transistor
 (SQUIPT)~\cite{giazotto_superconducting_2010}.

 In this design, robust phase-biasing performance is ensured by the
 adoption of a $150$-nm-thick aluminum loop of micrometric
 size, a configuration that proved to be effective in SQUIPTs based on
 normal-metal weak links of comparable
 size~\cite{ronzani_highly_2014,dambrosio_normal_2015}.
 Moreover, the pronounced geometric contrast in the cross section of the ring with
 respect to the superconducting wire minimizes depairing effects induced by
 supercurrent concentration at the interfaces between the wire and the thicker
 superconducting ring~\cite{vijay_approaching_2010}.

The modulation of the current-vs-voltage $I(V_b)$ characteristics of 
a typical (wire length $L=160\,\mathrm{nm}$) normal-metal tunnel probe device
(tunnel resistance $R_T \approx 150\,\mathrm{k\Omega}$) as a function of the
applied magnetic flux $\Phi$ is presented in Fig.~\ref{fig:nprobe}.
At base temperature ($T=20\,\mathrm{mK}$) increasing the magnetic flux bias from
$\Phi = 0$ to $\Phi=\Phi_0/2$ results in a 65\% suppression of the energy gap in
the quasiparticle DOS compared to its zero-field value (panels a,e). 
Notably, the low-temperature differential conductance
characteristics (see Fig.~\ref{fig:nprobe}e) recorded for $\Phi/\Phi_0 \lesssim 0.25$
are compatible with data reported for specimens in the constant phase gradient
regime~\cite{anthore_density_2003}. 
This could be expected since in this flux range, the CPR is essentially linear.
However, for $\Phi/\Phi_0 \approx 0.5$, a peculiar concentration of
quasiparticle states at the edges of the residual energy gap can be inferred
from the experimental data.
The latter feature, absent in short phase-biased normal-metal
wires~\cite{dambrosio_normal_2015}, appears reproducibly between different
samples, provided a sufficient phase difference is applied to the short
superconducting wire.
By increasing the temperature ($T=300,\,500,\,700\, \mathrm{mK}$, panels b-d) the 
$I(V_b)$ curves show evidence of the progressive suppression of the residual
energy gap at $\Phi = \Phi_0/2$. The magnetic modulation of the differential conductance
recorded at $T=650\,\mathrm{mK}$ (panel f) shows the transition between a
superconductor/insulator/normal-metal-like response at zero field to an almost
ohmic response at $\Phi=\Phi_0/2$.

These observations can be understood by considering the
temperature response~\cite{likharev_superconducting_1979}
of the CPR of a weak link based on a superconducting wire in
contact with rigid superconducting electrodes. 
The physical observables of the latter have been calculated in
the quasiclassical framework by solving the Usadel equations self-consistently
with the pairing amplitude profile.
Figure~\ref{fig:theo} shows a synopsis of modeled physical quantities obtained
from a parameter set chosen in accordance with experimental data
(refer to the Methods section for details).

As the temperature increases, the
current-to-phase relation of the weak link progressively shifts from a
multi-valued $I_s(\Phi)$ at low-temperature [characterized by
metastable $\Phi=\Phi_0 (n+1/2)$ nodes] to a
single-valued CPR functional form reached at $T=700\, \mathrm{mK}$. 
In the latter regime the amplitude of the pairing potential in
the centre of the wire can be completely suppressed by applying a phase
difference equal to $\pi$ (panel c). 

Simultaneously, the differential conductance as a function of voltage
bias, applied magnetic flux and temperature can be computed from the
corresponding DOS. In 
Figure~\ref{fig:theo}d--g, the calculated differential conductance maps
 are juxtaposed for comparison with data
measured at different temperatures. 
The striking correspondence obtained corroborates the physical
interpretation of complete pair potential suppression in the superconducting
wire for $\Phi/\Phi_0 = 0.5$.

\begin{figure}[h!]
	\centering
	\includegraphics{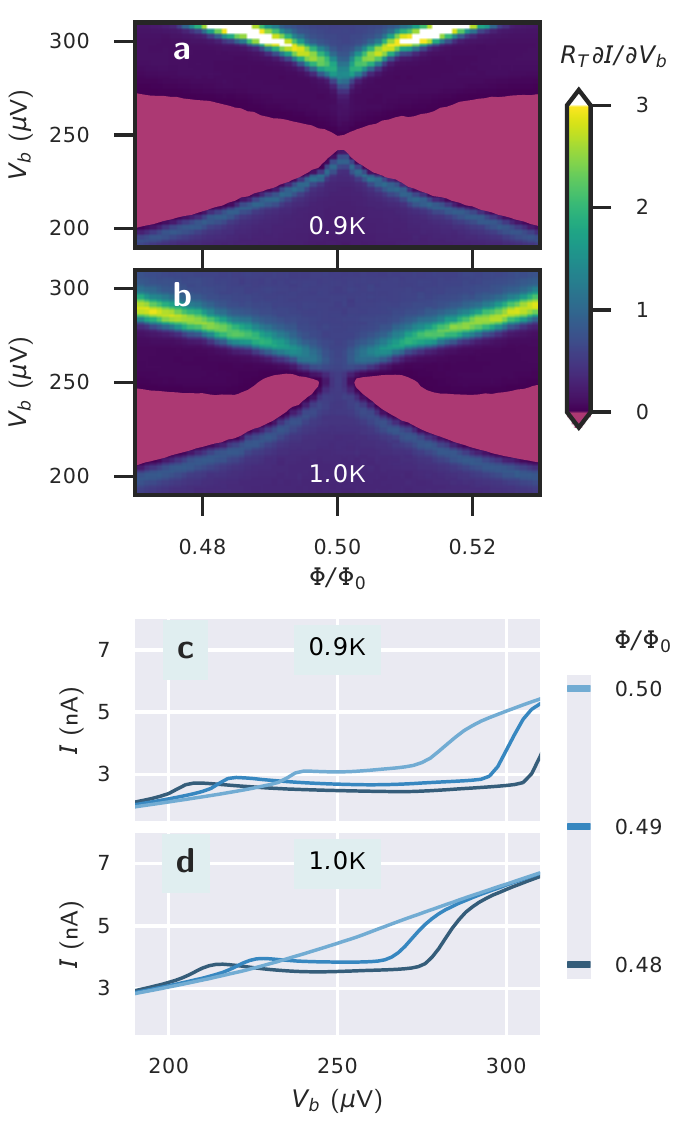}
\caption{Superconducting tunnel probe spectroscopy of the phase-driven collapse
of the superconducting gap in the Al nanowire.
\textbf{a, b}: Normalized differential conductance as a function of the applied
	magnetic flux and voltage bias, 
	recorded at lattice temperature $T = 0.9,\, 1.0\, \mathrm{K}$,
	respectively. The region showing negative differential conductance is 
	indicated in magenta.
	\textbf{c, d}: Current-vs-voltage characteristic curves recorded for
	$\Phi/\Phi_0 = 0.48,\, 0.49,\,0.5$ and for lattice temperature
	$T = 0.9,\, 1.0\, \mathrm{K}$. 
	}
\label{fig:sprobe}
\end{figure}

This above interpretation is further confirmed by observations focused on the transition
from the unstable to stable $\pi$ phase bias regime in devices equipped with a
superconducting tunnel electrode. The latter, realized by a
$15$-nm-thick oxidized aluminum film, features a BCS-like DOS 
characterized by a sizeable superconducting gap 
$\Delta_{pr} \approx 250\,\mathrm{\mu eV}$, typical of thin Al films.
As a consequence, the spectroscopic sampling of the DOS
in the phase-biased wire does not suffer from the loss of energy resolution due to thermal
broadening typical of normal-metal probes.
In particular, this setup allows for a direct estimate
of the energy gap $\varepsilon_g(\Phi)$ in the probed DOS.
At finite temperature, the latter quantity can be derived~\cite{meschke_tunnel_2011} from
the difference between voltage bias values relative to the direct and thermally-activated
conductance peaks (found, respectively, at $e\, V_b = \Delta_{pr} \pm
\varepsilon_g$, similarly to the well-known case of quasiparticles tunneling
between different superconductors at nonzero temperature~\cite{tinkham_introduction_1996}).

Panels a,b in Figure~\ref{fig:sprobe} show normalized differential conductance
maps recorded for a representative device characterized by a $L=210\,\mathrm{nm}$
superconducting wire in contact with a $R_T = 15\,\mathrm{k\Omega}$
superconducting tunnel electrode. The mapping is focused on voltage bias values
corresponding to the superconducting gap in the probe ($e\,V_b \approx
\Delta_{pr}$) and with coupled magnetic flux applied in a minute range
centered around $\Phi_0/2$. 
By inspecting the flux modulation of the direct ($e\,V_b > \Delta_{pr}$)
and thermally-activated ($e\,V_b < \Delta_{pr}$)
conductance peaks, an incomplete suppression of $2\, \varepsilon_g \simeq
40\,\mathrm{\mu eV}$ can be inferred
from data recorded at $T=0.9\mathrm{K}$ (panel a). 
Within a $100\,\mathrm{mK}$ temperature increase, we observe the merging of the
direct and thermally-activated peaks at $\Phi/\Phi_0 = 0.5$ and $e V_b = \Delta_{pr}$,
the direct evidence of the full suppression of the energy gap in the probed
quasiparticle DOS. Equivalently, the latter is associated with a smooth monotonic $I(V_b)$
characteristic curve at $T=1\mathrm{K}$ for $\Phi/\Phi_0 = 0.5$ (panel d), whearas the
corresponding curve at $T=0.9\mathrm{K}$ (panel c) displays a $\simeq
40\,\mathrm{\mu V}$-wide plateau.

We interpret these observations as the confirmation that the increase in temperature
has driven the CPR of the weak link to the single-valued regime, leading to a
complete collapse of the amplitude of the pairing potential in the center of the
wire for $\Phi/\Phi_0 = 0.5$.
In this case, the temperature value for this transition is
higher than for the normal-probe interferometer (Figures~\ref{fig:nprobe}
and~\ref{fig:theo}) as expected from the difference in the
respective lengths of the wires, in agreement with the theory.
Notably, while the value $T=1\mathrm{K}$ is arguably sizeable, is also 
significantly smaller than the critical temperature ($T_{c,w} = 1.4\mathrm{K}$,
see Methods) of the 25-nm-thick Al layer the wire consists of.

\begin{figure}[h!]
	\centering
	\includegraphics{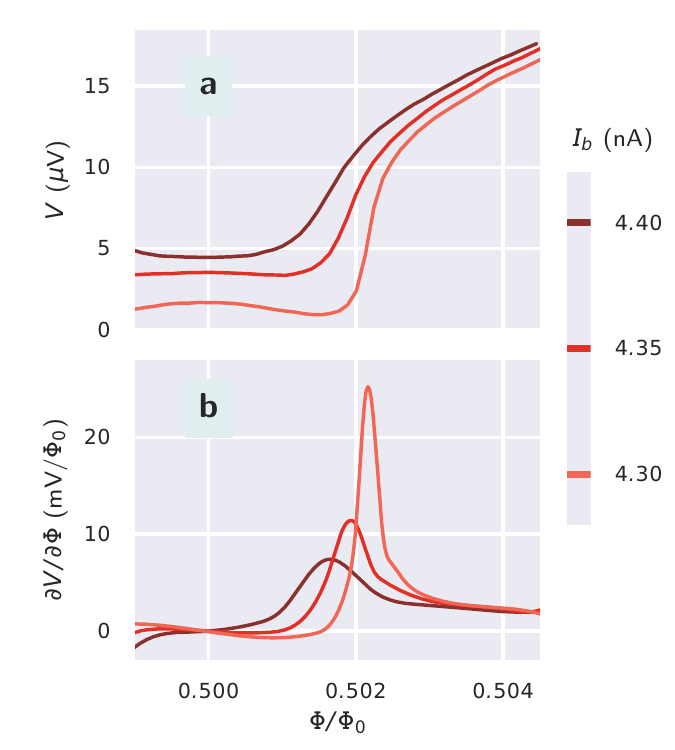}
\caption{Magneto-electric response of a typical superconducting tunnel probe device.
\textbf{a, b}: Respectively, voltage response (arbitrary offset)
	and corresponding flux-to-voltage transfer
	function for $\Phi \approx \Phi_0/2$,
	recorded at temperature $T= 1.0\, \mathrm{K}$. 
	The device is here operated under
	fixed current bias values $I_b = 4.30,\, 4.35,\, 4.40\, \mathrm{nA}$.
	}
\label{fig:smag}
\end{figure}

The steep character of the magnetic flux dependence of the pairing potential
suppression suggests to exploit these devices for
highly-sensitive magnetometry applications. 
Inspection of voltage traces recorded at $T=1\mathrm{K}$ under constant
current bias $I_b$ from the representative superconducting-probe device
(Figure~\ref{fig:smag}, panel a) reveals abrupt but continuous voltage
response in a minute magnetic flux range close to $\Phi/\Phi_0 = 0.5$. 
Here, the different traces indicate $I_b \in \left[ 4.3,\, 4.4\right]
\,\mathrm{nA}$, a range corresponding to the quasiparticle current measured in
the voltage-biased setup, with $V_b \approx \Delta_{pr}/e = 250\,\mathrm{\mu V}$.
The corresponding flux-to-voltage responsivity characteristics (panel
b) obtain values as large as $ \sim 27\, \mathrm{mV}/\Phi_0$, which are unparalleled in this class of 
devices~\cite{ronzani_highly_2014,dambrosio_normal_2015}.
In a simple DC-biased setup,
magnetic flux resolution figures as low as $
250\,\mathrm{n\Phi_0/\sqrt{Hz}}$, limited by shot noise intrinsic
to the tunnel-probe readout, have been measured in the $0.1 - 100\,
\mathrm{Hz}$ range (see Supplementary material).

In summary, we have presented a robust and reproducible means of suppressing
the order parameter amplitude of the Cooper condensate inside a
nanosized superconductor. Our observations are consistent with
established theory.
Reaching the complete flux modulation of the energy gap in the 
DOS inside the superconducting wire marks the
transition between the \emph{intrinsic}-like and
\emph{Josephson}-like regime. 
The corresponding temperature-dependent modulation of the CPR from a
multi-valued locus to a proper single-valued form entails, at the cross-over,
a strong dependence of the physical observables on the applied magnetic flux
for $\Phi \approx \Phi_0/2$. Yet, this property has been applied for the
realization of ultra-sensitive small-area quantum magnetometers.
More generally, this work provides experimental coverage of the
fundamental consequences of the self-consistency requirement on the order paramenter
in spatially inhomogeneous superconductors.
Finally, our design might be suitable as a testbed for the generation and investigation of
a spatially-locked ``phase-slip center''.
Another exciting perspective is to investigate whether the multivalued regions
of the CPR are reachable in pulsed phase experiments. 

\section{Acknowledgements}
We acknowledge funding from the European Research Council under the European
Unions Seventh Framework Program (FP7/2007-2013)/ERC Grant agreement
No.~615187-COMANCHE, the Italian Ministry of Education, University and Research
(MIUR) through the program FIRB-RBFR1379UX and the Tuscany Region through the
CNR joint project ``PROXMAG''. 
C.A. acknowledges discussions with H.
Pothier, P. Joyez and S. Bergeret at the root of the project.

\section{Methods}
\subsection{Fabrication protocols and experimental set-up}
The devices have been fabricated by directional metallic thin film deposition
through a self-aligned suspended mask. 
The latter is obtained by electron beam
lithography on a positive tone resist bilayer (1000 nm copolymer / 100 nm
poly-methyl metacrylate) spun on an oxidized Si wafer. 
After the development of the resist bilayer, the substrate is loaded in a
ultra-high vacuum (base pressure $\approx 10^{-9}$ Torr)  electron beam
evaporator, where an initial 15-nm-thick layer of 5N-Al/Al$_{0.98}$Mn$_{0.02}$ is
deposited at $40^\circ$ to realize a superconducting/normal-metal probe
electrode, respectively. A subsequent controlled exposition (300 s) to a pure
O$_2$ atmosphere with pressure values in the $10 \div 100$ mTorr range yields
tunnel junctions having specific resistance in the $20 \div 200 \,
\Omega\,\mathrm{\mu m}^2$ range.  
The superconducting interferometer is evaporated on top of the tunnel electrode as a
25-nm-thick Al layer at $20^\circ$, followed by a 150-nm-thick Al layer at
normal incidence.

After chemical lift-off in acetone followed by rinsing in isopropylic alcohol,
the samples are inspected in a scanning-electrode microscope.
Suitable devices are wire-bonded on a 24 pin dual-in-line ceramic chip carrier,
which is then loaded in a filtered $^3$He/$^4$He dilution refrigerator (Oxford
Instruments, mod. Triton 200) for the 
magneto-electric characterization down to $20\, \mathrm{mK}$. 
In the latter, the magnetic field is applied orthogonal to the substrate by a
custom-built superconducting electromagnet driven by a low-noise
programmable current source (Keythley model 2600). 

Low-noise current / voltage bias
sources (a lithium battery in series with a $200\, \mathrm{M\Omega}$ impedance and a
Yokogawa GS200, respectively) are coupled with room-temperature voltage / current
preamplifiers (NF Corporation LI-75A and DL Instruments 1201, respectively)
to realize the DC readout. For normal-metal tunnel electrode devices the
differential conductance is recorded with lock-in amplification (Stanford
research model 830) with $4\mu\mathrm{V}$ voltage excitation (rms).

\subsection{Theoretical model}
We model the properties of the superconducting wire embedded in the
ring by the quasiclassical Green function
model~\cite{belzig_quasiclassical_1999},
assuming a quasi-1D geometry. The spectral angle $\theta$ and phase
$\chi$ are solved from the Usadel equations~\cite{virtanen_spectral_2016}
\begin{gather}
  \label{eq:usadel}
\hbar{}D\partial_x^2\theta = -2iE\sinh\theta + \frac{D(\partial_x\chi)^2}{2}
	\sinh2\theta \\\notag\qquad + 2i|\Delta|\cos(\phi-\chi)\cosh\theta \,,
	\\ \hbar{}D\partial_x\cdot(\partial_x\chi \sinh^2\theta) =
	-2i|\Delta|\sin(\chi-\phi)\sinh\theta \,,
\end{gather}
where $D$ is the diffusion constant.  We assume the contacts between
the wire and the ring have nonzero resistance, and model them with the
Kupriyanov--Lukichev boundary conditions~\cite{kupriyanov_influence_1988}
\begin{align}
  \label{eq:kl} \mp{}r\sinh\theta\partial_x\chi &=
	\sin(\chi-\chi_{\mp})\sinh\theta_{\mp} \,, \\ \mp{}r\partial_x\theta &=
	\sinh\theta\cosh\theta_{\mp}
	-\cos(\chi-\chi_{\mp})\cosh\theta\sinh\theta_{\mp} \,,
\end{align}
at the left ($-$) and right ($+$) contacts.  The parameter $r=R_{I}/r_w$ is the
ratio between the interface resistance $R_I$ and the resistance per length $r_w$
of the wire.  The values $\theta_\pm=\artanh(|\Delta_r(T)|/E)$, $\chi_-=0$,
$\chi_+=\varphi$ are the values of the angles inside the superconducting ring.
Here, the $\Delta_r(T)$ value of order parameter in the ring is assumed to be
BCS-like, determined by the critical temperature $T_{c,r}$.
We neglect proximity effect in the ring, and assume it is unperturbed by
the weak link.

The normalized DOS $N(E,x)=\Re\cosh\theta(E,x)$ and supercurrent
$I_s=A\sigma\int_{-\infty}^\infty\dd{E}\Im[\partial_x\chi\sinh^2\theta]\tanh\frac{E}{2
k_B T}$ are obtained from the solutions.  The order parameter $\Delta=|\Delta|e^{i\phi}$
is obtained from the self-consistency relation,
\begin{align}
	\label{eq:selfcons} |\Delta| \ln\frac{T}{T_{c,w}} = 2\pi i
	  k_B T\sum_{\omega_n>0}\left[e^{i(\chi-\phi)}\sinh\theta -
	  \frac{|\Delta|}{E}\right]_{E=i \omega_n} \,,
\end{align}
where $T_{c,w}$ is the critical temperature of the wire, and $\omega_n = \pi k_B T
(2n +1 )$ are the discrete Matsubara energies.
Self-consistent determination is required for current conservation.

The solution to Eq.~\eqref{eq:selfcons} becomes multivalued when
$L\gtrsim{}\xi_0$. To trace the solution branch $(\varphi,\Delta(x))$, in this
case, we apply the pseudo-arclength continuation
method~\cite{keller_numerical_1977}. The
method generates the next point $(\varphi_{k+1},\Delta_{k+1}(x))$ based on
previous solutions, by requiring that $\Delta_{k+1}(x)$ satisfies
Eq.~\eqref{eq:selfcons}, and that the phase difference $\varphi_{k+1}$, which is
also considered unknown, satisfies a pseudo-arclength condition. This condition
is an approximate form for requiring the distance from the previous solution,
$s=\eta|\varphi_{k+1}-\varphi_k|^2+(2-\eta)||\Delta_{k+1}-\Delta_k||_2^2$ be
constant $s\approx{}s_0$, where $0<\eta<2$ is the weight factor and $s_0$ the
step size. 

In the comparison with experimental data, the relation between applied magnetic
flux $\Phi$ and phase difference $\varphi$ is assumed non-linear, with
$\Phi/\Phi_0 = \varphi/2\pi + \beta_L I_s(\varphi)/I_c$, where $I_c$ is the
low-temperature value of the critical current and $\beta_L = \mathcal{L} I_c/
\Phi_0$ accounts for nonzero ring inductance $\mathcal{L}$. 
The thermodynamically stable solution is determined as the one that
minimizes the free energy
$F(\varphi)=\frac{\Phi_0}{2\pi}\int_0^\varphi\dd{\varphi'}I_s(\varphi') +
\mathcal{L}I_s^2(\varphi)/2$.
The normalized differential conductance $R_T \partial I / \partial V_b$ is
calculated from the probe quasiparticle current
$ I(V_b, \Phi, T) = 1/(e R_T) \int dE\, \hat{N}(E, \Phi, T) [ f_0(E-eV_b) -
f_0(E)] $, where $f_0(E) = [1+\exp(E/k_B T)]^{-1}$ is the Fermi-Dirac distribution
function at temperature $T$ and $\hat{N}(E, \Phi, T) = 2/L \int_{L/4}^{3L/4}
N(E, x)_{T, \Phi}\, dx$ is the wire DOS
averaged over the typical physical length sampled by the probe.

The datasets in Figure~\ref{fig:theo} have been generated assuming the critical
temperature values $T_{c,w}=1.4\mathrm{K}$ for the thin wire
and the bulk value $T_{c,r}=1.25\mathrm{K}$ for the thick ring.
The modeled diffusive weak link has normalized length
$L/\xi_0 = 1.7$, corresponding to $\xi_0 = \sqrt{\hbar D / \Delta_0} =
95\,\mathrm{nm}$ for a physical length $L=160\,\mathrm{nm}$.
The interfaces of the wire are modeled with a non-ideality coefficient $r=0.75$.
The values of the set of mutually-independent parameters $T_{c,w},\, \xi_0,\, r$
are chosen on the basis of optimal reproduction of the differential conductance
characteristic curve recorded for null magnetic field at $T=20\,\mathrm{mK}$.
Notably, data recorded at higher temperatures is also satisfactorily reproduced.
Optimal agreement with the observed flux modulation at all temperatures
is obtained by letting
$\beta_L = 0.03$, consistent with $I_c \approx 18 \,\mathrm{\mu A}$ deduced from
the former parameters while assuming $\mathcal{L} = 3.5\, \mathrm{pH}$ (numerical estimate
of the inductance of the superconducting loop including both geometric and
kinetic contributions, with magnetic penetration depth $\lambda_\bot \approx
60\,\mathrm{nm}$.  FastHenry version 3.0wr by S.~R.~Whiteley, available from
\url{http://wrcad.com}).

\end{document}


\selectlanguage{english}
\title{Phase-driven collapse of the Cooper condensate in a nanosized
superconductor --- Supplementary Information}

\maketitle
\section*{Magnetometric performance}
In the main article's body it has been shown that superconducting-wire SQUIPT
magnetometers exhibit very pronounced flux-to-voltage responsivity figures when
their current-phase relation transitions from multi-valued to single-valued
regime. Here we expand on this application by assessing in detail the magnetometric
properties of a complete setup based on this type of device, as illustrated in
Figure~1. Notably, the analysis of the cross-correlation between the
output signals of two parallel amplification chains connected to the same
device allows to distinguish amplifier-limited magnetic flux resolution
performance from noise sources intrinsic to the readout scheme.

Figure~2 shows a summary of the voltage response of the representative
superconducting probe device for $\Phi\approx \Phi_0/2$, measured at temperature
$T=1\mathrm{K}$ under fixed current $I_b=4.35\, \mathrm{nA}$.
The noise characteristic of the readout/amplifier system can be assessed by
tuning the applied magnetic flux to $\Phi = \Phi_0/2$, where the voltage
response is null to the first order in $\Phi$. In this configuration (panel
\textbf{d}) the power spectral density profiles of the individual preamplifiers
(green/blue traces) are only barely higher than their nominal datasheet values,
whereas the cross-spectral density (black trace) converges to a profile
$\mathcal{X}_{12}(f)$ consistent with the following model (gray shade):
\begin{equation}
\mathcal{X}_{12}(f)= \sqrt{v_a^2 + R_{d}^2 \, \frac{ 2 e I_b }{\left| 1+2
	\pi \mathrm{i} f R_d \, \mathcal{C} \right|^2 }} \quad ,
	\label{eqn:model}
\end{equation}
where $v_a = 0.7\,\mathrm{nV/\sqrt{Hz}}$ is a white noise background, $e$ is the
elementary charge, $R_d = \partial V / \partial I_b$ is the differential
resistance and $\mathcal{C} = 26\, \mathrm{nF}$ is the effective shunt capacitance
consistent with the noise roll-off observed for $f \approx 100 \,\mathrm{Hz}$.
This simple model, based on the quadrature summation of RC-filtered tunnel
shot noise with an amplifier-dependent white noise background is sufficient to
describe $\mathcal{X}_{12}$ data recorded for $0.3 < f < 300\, \mathrm{Hz}$.

In Figure~2, a comparison between panel \textbf{b} and \textbf{c} shows that the peak in
the flux-to-voltage transfer function is associated with a corresponding
peak in the differential resistance of the device ($\Phi/\Phi_0=0.502$). 
On the other hand, the latter is suppressed with $\Phi/\Phi_0 > 0.5025$, while
the flux-to-voltage transfer function maintains appreciable levels.
Panel \textbf{e} shows indeed that at the working point associated with maximal
responsivity, the power spectral density of the individual amplifiers is
basically indistinguishable from the cross-correlated spectrum, \textit{i.e.},
the $v_a$ term is negligible in Equation~\ref{eqn:model}. Here, the high value of
$R_d$ induces both a reduction in the available bandwidth and an increase of the
shot-noise contribution to the observed voltage spectral density.
By contrast panel \textbf{f}, corresponding to a working point
($\Phi/\Phi_0=0.5025$) characterized by a significantly lower value of $R_d$,
demonstrates a $\mathcal{X}_{12}(f)$ profile characterized not only by wider
available RC bandwith, but also slightly better signal to noise ratio (as
suggested by the shape of the $1.3\,\mathrm{Hz}$ signal peak). 
This difference in performance can be more quantitatively appreciated by renormalizing
the spectral density profiles from voltage to magnetic flux units by means of
the flux-to-voltage transfer function values. The results are shown in Figure~3.
A summary of the performance figures for the different working points is
reported in Table~I.

\setlength{\tabcolsep}{8pt}
\begin{table}
	\begin{tabular}{r@{.}l r@{.}l r@{.}l c c r}
	\multicolumn{2}{c}{$\Phi$} & \multicolumn{2}{c}{$\partial V/\partial\Phi$} 
		& \multicolumn{2}{c}{$R_d$} & \multicolumn{2}{c}{noise floor} & bandwidth \\
		\multicolumn{2}{c}{$\Phi_0$} & \multicolumn{2}{c}{mV/$\Phi_0$} 
		& \multicolumn{2}{c}{k$\Omega$} & $\mathrm{nV/\sqrt{Hz}}$ &
		$\mathrm{n\Phi_0/\sqrt{Hz}}$ & $\mathrm{Hz}$\\
	\hline
		0&5 & 0 &0 & 36&7 & 1.5 & --- & 167 \\
		0&502 & 11&3 & 117& & 4.4 & 390 & 52 \\
		0&5025 & 4&6 & 26&2 & 1.2 & 260 & 234

\end{tabular}
	\caption{Summary of magnetometric figures for the working points in Figure~2.}
\end{table}

\begin{figure}[p]
\includegraphics{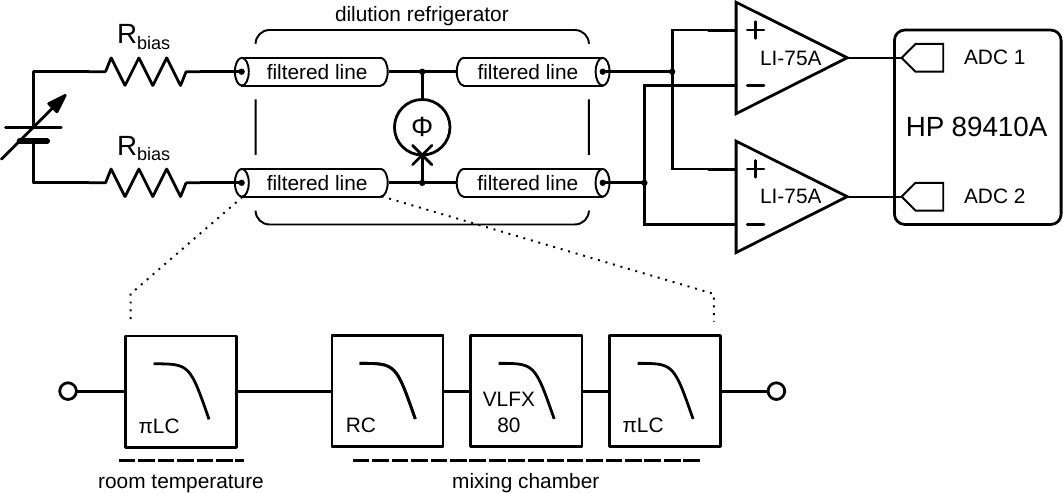}
	\caption{Functional schematic of the setup adopted for the characterization of
	the noise-equivalent flux level of SQUIPT
	magnetometers. The voltage response of the latter is probed in a 4-wire
	configuration where the current bias is provided by an adjustable
	battery-powered voltage source in series with bias resistors $R_{bias}=100\, \mathrm{M}\Omega$.
	The device is housed in a copper enclosure thermally anchored to the
	mixing chamber of a helium dilution refrigerator. The latter is equipped with
	identical electrical DC lines filtered as follows: room-temperature
	LC($\pi$) lowpass filter (Oxley FLT/P/5000), two-pole RC lowpass
	(room-temperature values: $R=500\,\Omega$, $C=47\, \mathrm{nF}$)
	at the mixing chamber
	followed by high-rejection multistage lowpass (Minicircuits VLFX-80) in
	series with a LC($\pi$) filter (Oxley SLT/P/5000) at the sample's
	enclosure. 
	The magnetic flux $\Phi$ linked to the interferometer loop is applied
	by a superconducting magnet surrounding the sample stage and powered by
	a low-noise programmable current source (Keythley model 2600, not shown).
	The voltage difference across the device is amplified by two separate
	battery-powered room-temperature differential voltage amplifiers (NF
	Corporation model LI-75A), whose outputs are digitized independently as
	the inputs of a vector signal analyzer (HP model 89410A). The latter
	computes the power spectral density of each input channel as well as the
	cross-spectral density between them.
	}
\end{figure}

\begin{figure}[p]
\includegraphics{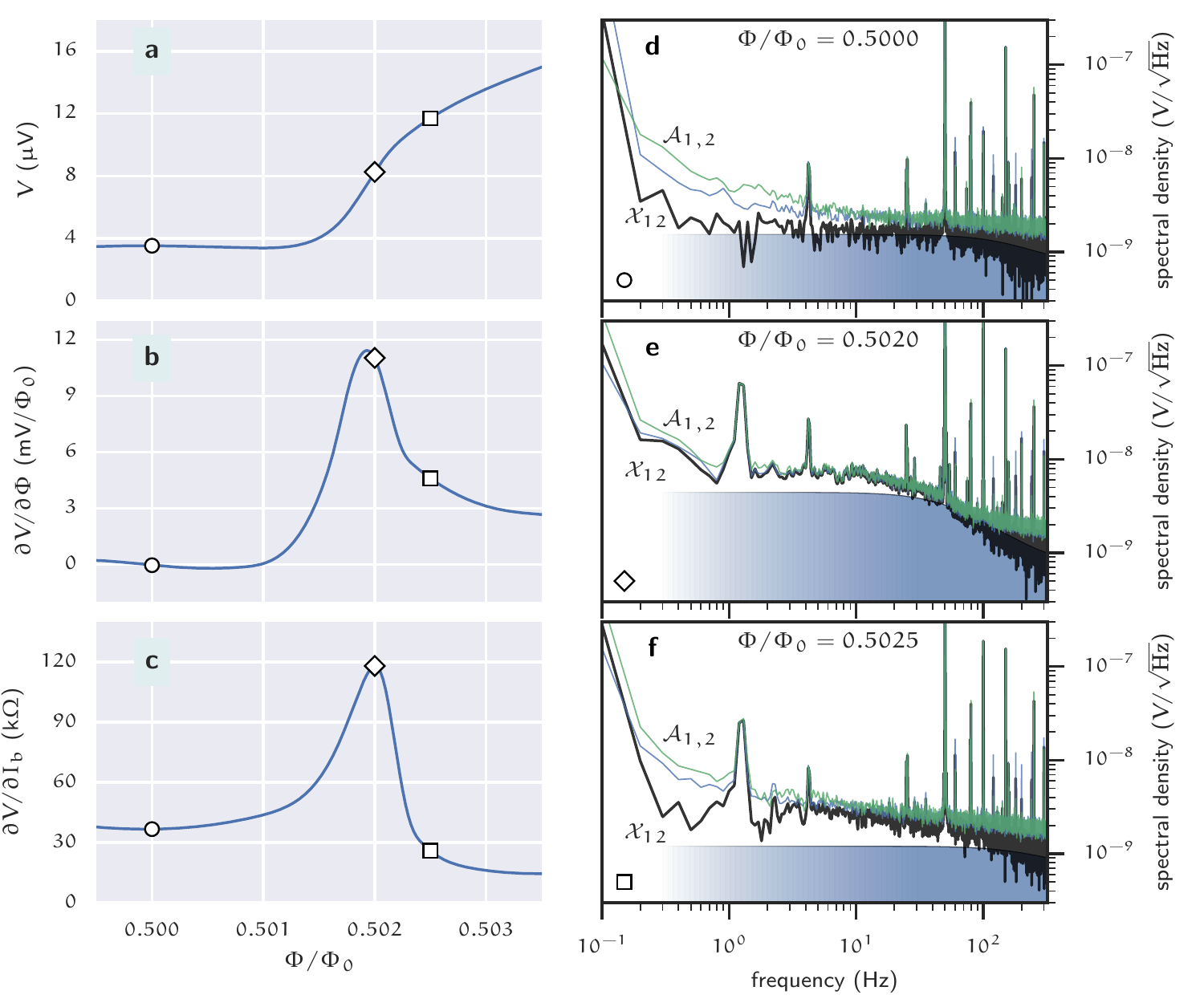}
	\caption{Summary of the magnetoelectric response figure of a typical
	superconducting-probe device recorded at temperature T=1K and current
	bias $I_{b} = 4.35\, \mathrm{nA}$. 
	The datasets shown here expand on the corresponding data presented in
	Fig.~4 in the main article's body.
	Panels \textbf{a}, \textbf{b} and \textbf{c} show as a function of the
	applied flux $\Phi$, respectively, the DC voltage, the flux-to-voltage
	transfer function (obtained by numerical differentiation) and the differential
	conductance (estimated by finite differences from the voltage traces
	recorded for $I_{b} = \, 4.3,\,4.35,\,4.4\, \mathrm{nA}$).
	The spectral characteristics of the amplified signal
	(green/blue traces: individual-channel power spectral density
	$\mathcal{A}_{1,2}$, black trace: cross-spectral density
	$\mathcal{X}_{12}$, gray shading: readout noise level estimate)
	are presented in panels \textbf{d-f} for three illustrative flux working
	points.  Panel \textbf{d} corresponds to zero first-order response
	($\Phi/\Phi_0=0.5$), while panels \textbf{e} and \textbf{f} are relative
	to the device being tuned for highest responsivity and best
	noise-equivalent flux resolution, respectively.
	}
\end{figure}

\begin{figure}[p]
\includegraphics{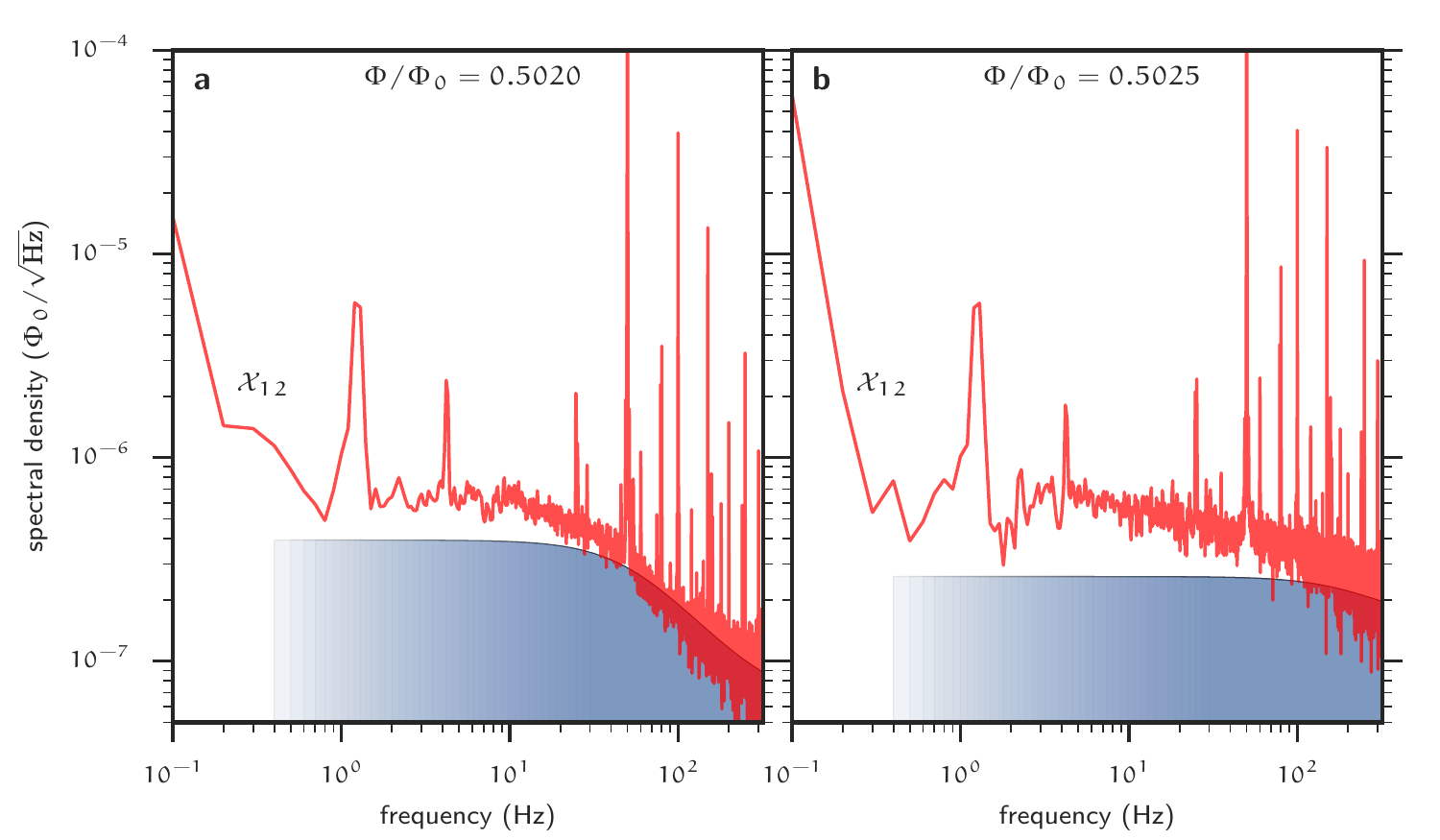}
	\caption{Cross-spectral density $\mathcal{X}_{12}$ (red trace) and readout noise
	estimate (gray shading) in magnetic flux units.  Panel \textbf{a} and
	\textbf{b} correspond, respectively, to the device tuned for highest
	responsivity ($\Phi/\Phi_0=0.502$) and for best noise-equivalent flux
	resolution ($\Phi/\Phi_0=0.5025$).
	}
\end{figure}